\documentclass[aps,pre,twocolumn,showpacs,superscriptaddress,groupedaddress,10pt]{revtex4-1}
\pdfoutput=1
\usepackage{hyperref}
\usepackage{epsfig}
\usepackage{amsmath}
\usepackage{graphicx}
\usepackage{wrapfig}
\usepackage{dcolumn}
\usepackage{bm}
\usepackage{amssymb}
\usepackage{titlesec}
\usepackage{mathtools}
\usepackage{relsize}
\usepackage{cleveref}
\usepackage[usenames,dvipsnames]{color}
\usepackage{ulem}
\usepackage[english]{babel}

\begin{document}
\title{Degree Dispersion Increases the Rate of Rare Events in Population Networks}
\author{Jason Hindes$^{1}$ and Michael Assaf$^{2}$} 
\affiliation{$^{1}$U.S. Naval Research Laboratory, Code 6792, Plasma Physics Division, Nonlinear Systems Dynamics Section, Washington, DC 20375, USA}
\affiliation{$^{2}$Racah Institute of Physics, Hebrew University of Jerusalem, Jerusalem 91904, Israel}

\begin{abstract}
There is great interest in predicting rare and extreme events in complex systems, and in particular, understanding the role of network topology in facilitating such events.
In this work, we show that degree dispersion -- the fact that the number of local connections in networks varies broadly -- increases the probability of large, rare fluctuations in population networks generically. We perform explicit calculations for two canonical and distinct classes of rare events: network extinction and switching. When the distance to threshold is held constant, and hence stochastic effects are fairly compared among networks, we show that there is a universal, exponential increase in the rate of rare events proportional to the variance of a network's degree distribution over its mean squared.
\end{abstract}
\maketitle

Systems containing a large, yet finite, population of interacting individuals or dynamical units often experience fluctuations due to the stochastic nature of agent interactions and local dynamics.
Most of the time, such systems reside in the vicinity of some attractor, undergoing small random excursions around it. Yet, occasionally a rare large fluctuation, on the order of the typical system size, may occur, which can lead to a transition to an absorbing state (a state that, once entered, cannot be left) or to the vicinity of another attractor. As a result, stochasticity can turn deterministically
stable attractors into \textit{metastable} states\cite{DykmanRev}. Examples of such extreme, rare events, which may be of key practical importance include population extinction \cite{Lande2003,Doering,Assaf2010,Meerson2013,Assaf2017}, switching in gene regulatory networks \cite{Assaf2011,MotterPRX2015,Biancalani2015,Bressloff2017}, the arrival of biomolecules at small cellular receptors \cite{Coombs2009}, and power-grid
destabilization \cite{Nesti,TimmePRE2017, HindesGrids2019}.

Usually, rare events in populations are considered within well-mixed or \textit{homogeneous} settings, e.g., where individuals interact with an equal number of neighbors. In this case, analytical treatment is possible using standard techniques \cite{Bressloff2017,Weber2017,Assaf2017}. On the other hand, it is known that in topologically \textit{heterogeneous} networks, e.g., where nodes have variable degree, the critical behavior can be dramatically affected \cite{Moore2000,Dorogovtsev2002,Sood2005,Sood2008}. Unfortunately, predicting rare events in degree-heterogenous networks is notoriously hard, due to high dimensionality and complex coupling between degrees of freedom. Though some progress has been made by applying semi-classical approximations to master equations governing stochastic dynamics in complex systems ~\cite{Assaf2012,HindesPRL2016,Sabsovich2017}, often, the resulting Hamilton equations are difficult to solve, as they require computing unstable trajectories in high-dimensional phase spaces \cite{WEPRE2002,Lindley2,Schwartz2,Nieddu}. Consequently, analyzing rare events in general networks has been mainly limited to near-bifurcation regimes, where dimensionality is reduced.


In this Letter we apply a novel perturbation scheme that allows us to predict a universal increase in the rate of rare events by exploiting the \textit{extent} of network heterogeneity, or degree dispersion. We find that  this increase is proportional  to the ratio of the variance of a network's degree distribution to its mean squared, or coefficient of variation (CV) squared, and is otherwise independent of topology. Our approach is shown analytically for two canonical examples of fluctuation-driven rare events: extinction of epidemics in the Susceptible-Infected-Susceptible (SIS) model on networks, and switching (or spontaneous magnetization flipping) in binary spin networks. 

\textit{Extinction in heterogenous networks: the SIS model.}
We begin by considering the SIS model of epidemics, which consists of two types of individuals: susceptibles (S) and infecteds (I)\cite{Keeling1}. A susceptible can get infected upon encountering an infected individual, $S+I\to I+I$, while an infected can recover and become susceptible again, $I\to S$. We first consider networks with only two degree classes, and then generalize to arbitrary degree distributions. We assume a network of $N\gg 1$ nodes, with $N/2$ nodes of degree $k_1\!\equiv\!k_{0}(1-\epsilon)$ and $N/2$ nodes of degree $k_2\!\equiv\!k_{0}(1+\epsilon)$. Each node represents a single individual which can be in either state. We assume the infection rate is $\lambda$ and the recovery rate is $1$.

Denoting by $n_i$ the number of degree-$k_i$ ($i=1,2$) infected nodes, and by $x_i\!=\!n_i/(N/2)$ the densities of degree-$k_i$ infected nodes, the probability for a given node to be connected to an infected node in a random network with this {\it bimodal} degree distribution is
$\Phi(n_1,n_2)\equiv\Phi(x_1,x_2)=(k_1 x_1+k_2 x_2)/(k_1+k_2)$. Thus, the infection rate (per individual) of a susceptible node of degree $k_i$ is $\lambda k_i (1-x_i)\Phi(x_1,x_2)$, while the recovery rate is simply $x_i$.

In order to make analytical progress, we assume that the average dynamics over an ensemble of uncorrelated random networks can be approximated by the following four (twice the number of degree classes) stochastic reactions, occurring in a well-mixed setting~\cite{Sood2005,Sood2008,Assaf2012,HindesPRL2016,Sabsovich2017}:
\begin{eqnarray}
&&n_1\xrightarrow{\lambda k_1(N/2-n_1)\Phi(x_1,x_2)}n_1+1,\;\;\;n_1\xrightarrow{n_1}n_1-1,\nonumber\\
&&n_2\xrightarrow{\lambda k_2(N/2-n_2)\Phi(x_1,x_2)}n_2+1,\;\;\;n_2\xrightarrow{n_2}n_2-1.
\end{eqnarray}
This formulation is equivalent to the so called {\it annealed network approximation} (ANA) \cite{Pastor}. However, an analogous argument can be developed for networks with empirical adjacency matrices in the limit of large spectral gaps \cite{HindesPRE2017}. In the latter case, the degree is replaced by the eigenvector centrality in all results below.

We are interested in quantifying how broadening a network's degree distribution affects the rate of extinction of infection by stochastic fluctuations. We focus on the case where the standard deviation of the degree distribution, $\sigma$, is sufficiently smaller than its mean $\langle k\rangle$, allowing for a rigorous perturbative treatment. For bimodal networks $\langle k\rangle\!\equiv\!k_{0}$, while $\sigma\!=\!\sqrt{\langle k^2 \rangle-\langle k \rangle^2}\!=\!k_{0}\epsilon$. Therefore, we assume henceforth that $\sigma\ll \langle k \rangle$, or $\epsilon\ll 1$.

The deterministic rate equations, describing the mean density of infected nodes with degrees $k_1$ and $k_2$, read
\begin{eqnarray}\label{RE}
\dot{x}_1&=&\lambda k_{0}(1-\epsilon)(1-x_1)\Phi(x_1,x_2)-x_1,\nonumber\\
\dot{x}_2&=&\lambda k_{0}(1+\epsilon)(1-x_2)\Phi(x_1,x_2)-x_2.
\end{eqnarray}
The critical value of $\lambda$, below which there is no long-lived endemic state, satisfies on random networks $\lambda_c\!\equiv\! \langle k\rangle/\langle k^2\rangle\!=\!1/[k_{0}(1+\epsilon^2)]\!\simeq\! (1-\epsilon^2)/k_{0}$ (given the ANA)~\cite{Pastor}. Thus, we write $\lambda\!=\!\Lambda\lambda_c$, where $\Lambda\geq 1$, and $\Lambda\!-\!1$ measures the distance to bifurcation, or threshold.

Rate equations (\ref{RE}) admit two positive fixed points. For $\epsilon\!\ll\! 1$, these become: $[x_1,x_2]\!=\![x_0(1-\epsilon/\Lambda),x_0(1+\epsilon/\Lambda)]$, which is stable, and $[x_1,x_2]\!=\![0,0]$, which is unstable, where $x_0\!=\!(\Lambda-1)/\Lambda$. A {\it transcritical bifurcation} occurs as $\Lambda$ passes the value of $1$.
While it gives some intuition, the deterministic picture ignores demographic noise emanating from the discreteness of individuals and stochasticity of the reactions. This noise, and the fact that the
extinct state $n_1 = n_2 = 0$ is absorbing, make the non-trivial stable fixed point in the language of the rate equations, metastable. Thus, the network ultimately goes extinct via a rare, large fluctuation~\cite{KamenevPRE2008,Assaf2010,Assaf2017,Clancy2018,Holme2018}.


Accounting for demographic noise, the master equation for $P_{n_1,n_2}(t)$: the probability to find at time $t$, $n_1$ and $n_2$ infected nodes on degrees $k_1$ and $k_2$, respectively, satisfies
\begin{eqnarray}\label{master}
\dot{P}_{n_1,n_2}(t)&=&\left[\lambda  k_{0}(1-\epsilon)(E_{n_1}^{-1}-1)(N/2-n_1)\Phi(n_1,n_2)\right.\nonumber\\
&+&\lambda  k_{0}(1+\epsilon)(E_{n_2}^{-1}-1)(N/2-n_2)\Phi(n_1,n_2)\nonumber\\
&+&\left.(E_{n_1}^{1}-1)n_1+(E_{n_2}^{1}-1)n_2\right]P_{n_1,n_2},
\end{eqnarray}
where $\lambda\!=\!\Lambda(1-\epsilon^2)/\!k_{0}$, and $E_n^jf(n)=f(n+j)$ is a step operator. Next, we assume that the network settles into a long-lived metastable state prior to extinction. This assumption is justified if $N$ is large, and the mean time to extinction (MTE), $T$, is very long (see below). This metastable state, which is described by a quasi-stationary distribution (QSD) about the stable fixed point, slowly decays in time at a rate which equals $1/T$, while simultaneously the extinction probability grows and reaches the value of $1$ at infinite time \cite{DykmanRev,Assaf2010}. We now plug the ansatz $P_{n_1,n_2}\simeq \pi_{n_1,n_2}e^{-t/T}$ into master equation~(\ref{master}), where $\pi_{n_1,n_2}$ is the QSD, and employ the WKB approximation for the QSD, $\pi_{n_1,n_2}\equiv \pi(x_1,x_2)\sim e^{-NS(x_1,x_2)}$, where $S(x_1,x_2)$ is the action function~\cite{DykmanRev}. In the leading  order  in $N\gg 1$ we arrive at a Hamilton-Jacobi equation $H(x_1,x_2,\partial_{x_1}S,\partial_{x_2}S)=0$, with Hamiltonian
\begin{eqnarray}\label{hamSIS}
&&H(x_1,p_1,x_2,p_2)=\frac{\lambda k_0}{2}\Phi(x_1,x_2)\left[(1\!-\!\epsilon)(1\!-\!x_1)(e^{p_1}\!-1)\right.\nonumber\\
&&\!+\!\left.(1\!+\!\epsilon)(1\!-\!x_2)(e^{p_2}\!-\!1)\right]\!+\!\frac{x_1}{2}(e^{-p_1}\!-\!1)\!+\!\frac{x_2}{2}(e^{-p_2}\!-\!1),
\end{eqnarray}
where $p_{i}/2\!=\!\partial_{x_i}S$ are normalized momenta. The Hamilton equations satisfy $\dot{x}_{i}/2\!=\!\partial_{p_i}H$ and $\dot{p}_{i}/2\!=\!-\partial_{x_i}H$. Once $S(\bold{x})$ is known, by solving Hamilton's equations, so is the MTE, which is proportional to $e^{NS(0,0)}$~\cite{KamenevPRE2008,Assaf2010,HindesPRL2016}.

\begin{figure}[!h]
	\includegraphics[scale=0.32]{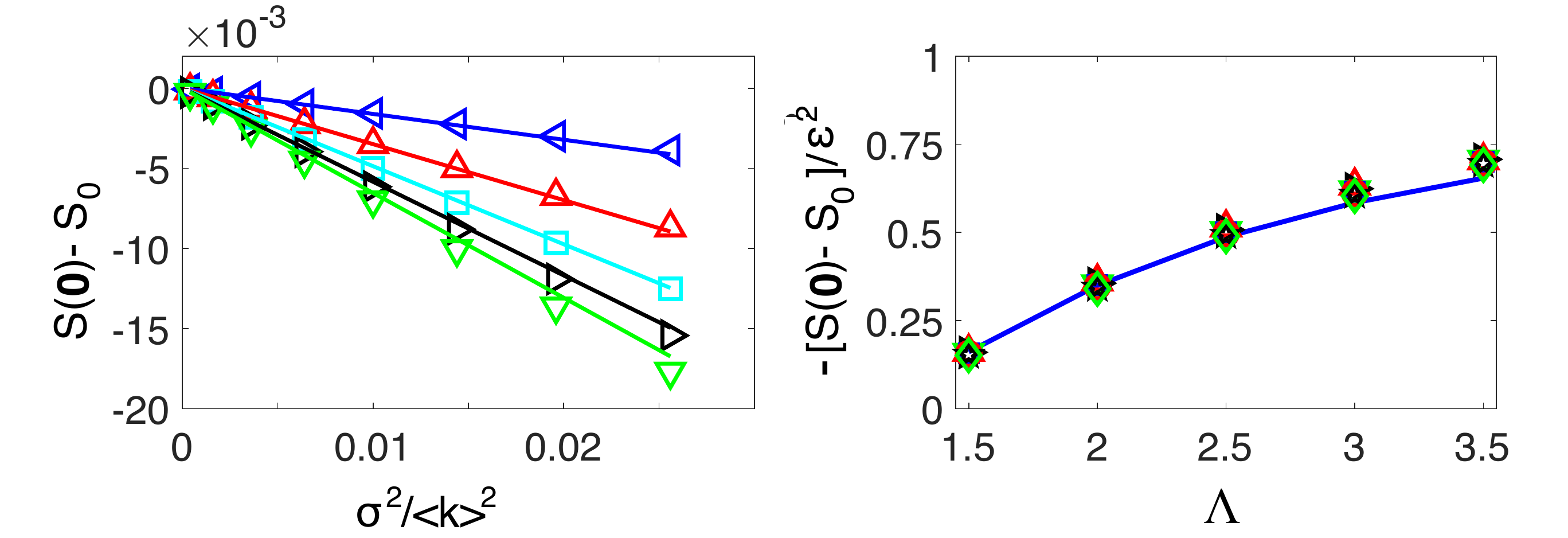}
	\caption{Left panel: $S(\mathbf{0})-S_0$ versus $\epsilon^2=\sigma^2/\langle k\rangle^2$ for bimodal networks. Symbols are numerical solutions of the Hamilton equations for $\Lambda=1.5,2,2.5,3,3.5$ (top to bottom), lines are the analytical results~(\ref{finalaction}). Right panel: $-[S(\mathbf{0})-S_0]/\epsilon^2$ versus $\Lambda$. Symbols are numerical solutions for $\epsilon=0.02-0.16$ (see left panel). The curve is the second of Eqs.~(\ref{finalaction}).}
	\label{fig:action}
\end{figure}

For convenience, let us define new variables $u=(x_1-x_2)/2$, $p_u=p_1-p_2$, $w=(x_1+x_2)/2$ and $p_w=p_1+p_2$. This transformation is canonical since the determinant of the Jacobian $\partial(\textbf{Q},\textbf{P})/\partial(\textbf{x},\textbf{p})=1$, where $\textbf{Q}=(u,w)$, $\textbf{P}=(p_u,p_w)$, $\textbf{x}=(x_1,x_2)$, and $\textbf{p}=(p_1,p_2)$. Using the new variables, the path to extinction connects between the fixed points $[w^*,u^*,0,0]$ and $[0,0,p_w^*,p_u^*]$, where
\begin{eqnarray}\label{fixedpoints}
w^*&=&x_0\left[1-(2/\Lambda)\epsilon^2\right],\;\;\;u^*=-(x_0/\Lambda)\epsilon,\\
p_w^*&=&-2\ln\Lambda+[x_0(3\Lambda+1)/\Lambda]\epsilon^2,\;\;\;p_u^*=2x_0\epsilon.\nonumber
\end{eqnarray}
Since the transformation of variables is canonical, the action along the path to extinction is given by~\cite{DykmanRev}
\begin{equation}\label{actionSIS}
S(\mathbf{0})=\frac{1}{2}\int p_1dx_1+\frac{1}{2}\int p_2dx_2=\frac{1}{2}\int p_wdw+\frac{1}{2}\int p_udu.
\end{equation}
Transforming to the new variables in Hamiltonian~(\ref{hamSIS}), and assuming $u$ and $p_u$ scale as ${\cal O}(\epsilon)$, we find the trajectories $p_w(w)$ and $p_u(u)$ up to ${\cal O}(\epsilon^2)$~\cite{SM}. The trajectories are then substituted into Eq.~(\ref{actionSIS}), which yields
\begin{eqnarray}\label{finalaction}
\hspace{-8.10mm}S(\mathbf{0})&=&S_0-f_{E}(\Lambda)\epsilon^2,\nonumber\\
\hspace{-8.10mm}f_{E}(\Lambda)&=&\left[(\Lambda\!-\!1)(1\!-\!12\Lambda\!+\!3\Lambda^2)\!+\!8\Lambda^2\ln\Lambda\right]/(4\Lambda^3),
\end{eqnarray}
where, $S_0=1/\Lambda+\ln\Lambda-1$ is the action for a degree-homogeneous network ($\epsilon=0$), and $f_{E}(\Lambda)>0$.
We have obtained an exponential increase in the rate of extinction due to network heterogeneity, which only depends on the CV of the network's degree distribution.
In Fig.~\ref{fig:action} we demonstrate that in the limit of $\epsilon\ll 1$ our analytical results~(\ref{finalaction}) agree well with numerical solutions of the Hamilton equations, obtained using the Iterative Action Minimization Method~\cite{Lindley2,SM}.

Given our analysis for bimodal networks, it is straightforward to generalize to arbitrary, symmetric degree distributions, first, and then to skewed distributions. Let us denote by $g(k)$ the node degree distribution. That is, if $N_k$ are the number of nodes of degree $k$ such that $\sum_k N_k=N$, we have $g(k)=N_k/N$. We assume that $g(k)$ is a symmetric distribution about the mean $k_0\!\equiv\!\langle k\rangle$, such that $g(k_0+i)=g(k_0-i)$ for $i=1,2,3,\dots$. Let us also assume our distribution has a bounded support such that $k_{\text{min}}=k_0-\Delta$ and $k_{\text{max}}=k_0+\Delta$, where $g(k<k_{\text{min}})=g(k>k_{\text{max}})=0$. We again denote by $n_k$ the number of infected individuals on degree-$k$ nodes, and by $x_k=[1/g(k)]n_k/N=n_k/N_k$ the fraction of such infected individuals. Writing down the master equation for $P_{\{n_k\}}$ -- the joint probability to find $(n_{k_{\text{min}}},\dots,n_{k_{\text{max}}})$ infected nodes of degree $k$, and using the above WKB formalism, $P(\textbf{x})\sim e^{-NS(\textbf{x})}$, where $\textbf{x}=(x_{k_{\text{min}}},\dots,x_{k_{\text{max}}})$, we arrive at a Hamiltonian equivalent to \cite{HindesPRL2016}. Denoting $g(k)p_{k}\!=\! \partial S\!/\partial{x_{x}}$, the action can be shown to satisfy~\cite{SM}
\begin{eqnarray}\label{genactSIS}
&&S(\mathbf{0})=\sum_{k=k_0-\Delta}^{k_0+\Delta} g(k) \!\int\! p_k dx_k=g(k_0)\!\int\!p_{k_0} dx_{k_0}\nonumber\\
&&+\sum_{j=1}^{\Delta} g(k_0-j)\!\int\! p_{k_0-j} dx_{k_0-j}+p_{k_0+j} dx_{k_0+j},
\end{eqnarray}
where we have used the symmetry of $g(k)$ about its mean $k_0$.
Now, since each {\it pair} of nodes $k_0\pm j$ for $j\in[1,\Delta]$ can be viewed as a bimodal network, using Eqs.~(\ref{actionSIS}) and (\ref{finalaction}), the action for such a bimodal network with degrees $k_0-j$ and $k_0+j$, satisfies:
$(1/2)\int p_{k_0-j} dx_{k_0-j}+p_{k_0+j} dx_{k_0+j}=S_0-f_{E}(\Lambda)\epsilon_j^2$, where $\epsilon_j\!=\!j/k_0$. Moreover, the node of rank $k_0$ can be viewed as a bimodal network with $\epsilon_j=0$, such that $\int p_{k_0} dx_{k_0}=S_0$. Therefore, using the fact that $\sum_k g(k)=1$ and that the variance of $g(k)$ satisfies $\sigma^2=\sum_k (k-k_0)^2 g(k)$, the action [Eq.~(\ref{genactSIS})] and MTE become:
\begin{eqnarray}\label{eq:ActionExtinction}
T\sim e^{NS(\mathbf{0})},\;\;\;S(\mathbf{0})=S_0-f_{E}(\Lambda)\,\sigma^{2}/\left<k\right>^{2}.
\end{eqnarray}

\begin{figure}
	\includegraphics[scale=0.27]{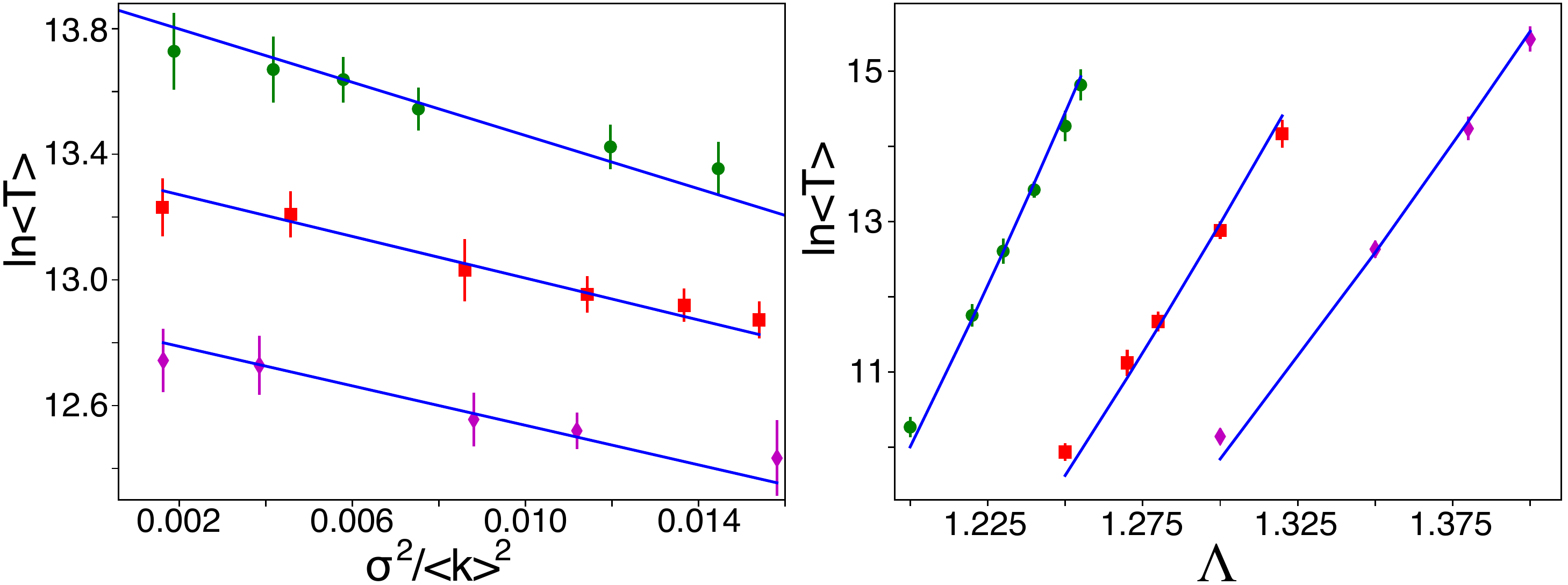}
	\caption{Left panel: MTE versus the degree dispersion for several networks; for each point, a mean time is computed from 200 stochastic realizations in a fixed network with a given degree distribution. This is repeated for 20 different network realizations with the same degree distribution and the same number of edges. The log of all such averages is then averaged. Error bars are given by the standard deviation of the latter. Results are shown for uniform (green, $\Lambda\!=\!1.16$, $N\!=\!1500$, $\left< k \right>\!=\!50$), Gaussian (red, $\Lambda\!=\!1.24$, $N\!=\!600$, $\left< k \right>\!=\!108.5$), and Gamma (magenta, $\Lambda\!=\!1.26$, $N\!=\!500$, $\left< k \right>\!=\!110.4$) distributions. Note that each distribution has one tunable parameter for the variance given a fixed $\left<k\right>$. Right panel: MTE versus the threshold parameter $\Lambda$. Results are shown for: Erd\H{o}s-R\'{e}nyi networks (green $N\!=\!600$, $\left< k \right>\!=\!160$, $\sigma\!/\!\left<k\right>\!=\!0.067$) and (magenta $N\!=\!300$, $\left< k \right>\!=\!120$, $\sigma\!/\!\left<k\right>\!=\!0.072$), and Gaussian distributions (red $N\!=\!400$, $\left< k \right>\!=\!110.4$, $\sigma\!/\!\left<k\right>\!=\!0.064$). Averages were computed in the same way as for (left).}
	\label{fig:actioneps}
\end{figure}

Equation (\ref{eq:ActionExtinction}) is the first of the main results in this work. Namely for any network, if the CV is small, $\sigma/\langle k\rangle\!\ll\! 1$, the logarithm of the MTE \textit{decreases linearly} with the square of the CV, compared to the degree-homogenous limit. This indicates that for large networks, for which $\sigma/\langle k\rangle\!\gg\! N^{-1/2}$, the extinction rate is exponentially increased when the population resides on a degree-heterogeneous network, compared with the homogenous case -- examples include human contact networks such as \cite{Salathe,Vespignani1}. Furthermore, while the pre-factor for the relative increase of the logarithm of the MTE, $f_{E}(\Lambda)$, is problem specific, it is \textit{independent} of the network topology, and is computed for any distance to threshold. Figure \ref{fig:actioneps} shows a comparison between Eq.~(\ref{eq:ActionExtinction}) and Monte-Carlo simulations for the MTE in several networks, demonstrating the agreement both in terms of $\sigma^2\!/\!\langle k\rangle^2$ and $\Lambda$.

Our analysis above required that the network degree distribution be symmetric and bounded. However, even for non-bounded \textit{asymmetric} distributions the MTE is still given by Eq.~(\ref{eq:ActionExtinction}), as long as such distributions are symmetric in the vicinity of their mean and their \textit{skewness} $\gamma_1$ is small. In fact, one can show that if these conditions are met, the errors contributed from neglected terms, outside of the symmetrical bulk, are negligible~\cite{SM}. This is demonstrated in Fig.~\ref{fig:actioneps} where we show that theoretical expression~(\ref{eq:ActionExtinction}) agrees well with numerics, also in the case of asymmetric Gamma distributions. Moreover, in the SM we show that our results even hold for power-law networks when the CV is not too large\cite{SM}.

\textit{Switching in heterogenous networks: the Spin model.}
Next, we consider a canonical binary spin system, where nodes are either (+) or  (-), instead of infected or susceptible, and make stochastic transitions according to a continuous-time Glauber dynamics \cite{Vespignani1,RednerBook}. Namely, if there is no spontaneous transition (analogous to spontaneous recovery in the SIS model), then each node $i$ flips spin at a rate proportional to $1/[1+\exp{\!\{\lambda\Delta E_{i}\}}]$, where $\Delta E_{i}$ is the change in the local pair-wise ferromagnetic energy for node $i$ to flip spin, and $\lambda$ is an inverse temperature. Here, the densities, $x_{k}$, are the {\it magnetization} of nodes with degree $k$: the fraction of degree-$k$ nodes with spin $(+)$ minus those with spin $(-)$. The master equation and Hamiltonian for $\bold{x}$ can be derived in precisely the same way as the SIS model above \cite{HindesSR2017}. The Hamiltonian reads
\begin{align}
\!\!H(\mathbf{x},\mathbf{p})=&\frac{1}{2}\sum_{k}g(k)\!\left[(1\!-\!x_{k})(e^{2p_{k}}\!-\!1)(1\!+\!e^{-2\lambda k\bar{x}})^{-1}\!\!\right.\nonumber \\
+&\left.(1\!+\!x_{k})(e^{-2p_{k}}\!-\!1)(1\!+\!e^{2\lambda k\bar{x}})^{-1}\right]\!,
\label{eq:spinHam}
\end{align}
where $\bar{x}\!=\!\sum_{k}\!kg(k)x_{k}/\!\left<k\right>$ is the degree-weighted mean magnetization, and $g(k)p_{k}\!=\!\partial S\!/\partial x_{k}$ are the momenta.

In contrast to the SIS model, the spin model exhibits three fixed points: $\mathbf{x}\!=\!\mathbf{x}^{*}$ and $\mathbf{x}\!=\!-\mathbf{x}^{*}$ which are stable, and $\mathbf{x}\!=\!\mathbf{0}$ which is unstable. The stable fixed points emerge at a {\it pitchfork bifurcation} when $\lambda=\lambda_{c}\!\equiv\!\left< k\right>\!/\!\left<k^{2}\right>$. As before, we may denote $\lambda\!=\!\Lambda\lambda_{c}$, where $\Lambda\!=\!1$ is the bifurcation threshold. In the spin model, demographic noise causes switching between $\mathbf{x}^{*}$ and $-\mathbf{x}^{*}$ \cite{ChenChaos}. In order to find the action for switching, we exploit the fact that there is detailed balance in the absence of spontaneous flipping (though this assumption can be relaxed without qualitatively changing our main result \cite{FN_DB}). As a consequence, the deterministic trajectory starting from the vicinity of the unstable point $\mathbf{0}$ and ending at the stable fixed point $\mathbf{x}^{*}$, coincides up to time reversal, with the fluctuational path from $\mathbf{x}^{*}$ to $\mathbf{0}$~\cite{DykmanRev}. Once at the unstable point $\mathbf{0}$, the network can switch to $-\mathbf{x}^{*}$ following its deterministic dynamics.

In order to find the switching path, we again use Hamilton's equations $g(k)\dot{x}_{k}\!=\!\partial H/\partial p_{k}$. The relevant trajectories $p_{k}(\mathbf{x})$ can be found by equating $-\dot{x}_{k}|_{\mathbf{p}=\mathbf{0}}=\dot{x}_{k}(\mathbf{p})$, where the former represents the deterministic trajectory. By doing so, the switching path satisfies~\cite{SM}
\begin{eqnarray}
p_k(\mathbf{x})=(1/2)\ln\left[(1+x_k)/(1-x_k)\right]-\lambda k\bar{x},\nonumber
\end{eqnarray}
and hence the action for switching, $S(\mathbf{0})\!=\!\sum_{k}\!g(k)\int_{x_{k}^{*}}^{0_{k}} p_{k}dx_{k}$, becomes
\begin{equation}
S(\mathbf{0})=\frac{\lambda\!\left<k\right>\!{\bar{x}^{*}}{^{2}}}{2}-\frac{1}{2}\!\sum_{k}g(k)\!\Bigg[\!\ln\!\big\{\!1-{x_{k}^{*}}^{2}\big\}+x_{k}^{*}\ln\!\bigg\{\!\frac{1+x_{k}^{*}}{1-x_{k}^{*}}\!\bigg\}\!\!\Bigg]\!.
\label{eq:FullAction}
\end{equation}

Following the same general approach as for the SIS model above, we write $k\!=\!k_{0}(1+\epsilon)$ where $\epsilon\equiv (k-k_0)/k_0$. For degree distributions with a small CV, $\sigma/k_0\ll 1$, we have $\lambda\!\approx\!\Lambda[1-\left<\epsilon^{2}\right>]/k_{0}$ and $\left<\epsilon^{2}\right>=\sigma^{2}/k_{0}^{2}$, as before. In order to evaluate Eq.~(\ref{eq:FullAction}) in the limit of $\left<|\epsilon|\right>\!\ll\!1$, we use the small-$\left<|\epsilon|\right>$ expansion of $x_{k}^*$ and $\bar{x}^*$, see~\cite{SM}, and keep terms up to order $\left<\epsilon^2\right>$. This procedure yields the action and mean switching time (MST)
\begin{eqnarray}
\label{eq:spinform}
\hspace{-8.10mm}T\sim e^{NS(\mathbf{0})};\;\;\;S(\mathbf{0})&=&S_{0}-f_{S}(\Lambda)\,\sigma^{2}/\left<k\right>^{2},\nonumber\\
\hspace{-8.10mm}f_{S}(\Lambda)&=&(\Lambda x_0^2/2)\left[1-\Lambda(1- x_0^2)\right],
\end{eqnarray}
where $S_{0}=-(1/2)\left[\ln\left(1-x_0^2\right)+\Lambda x_0^2\right]>0$, $x_{0}$ is the positive solution of $x_{0}\!=\!\tanh\{\Lambda x_{0}\}$, and $f_{S}(\Lambda)>0$.

\begin{figure}[!h]
	\includegraphics[scale=0.27]{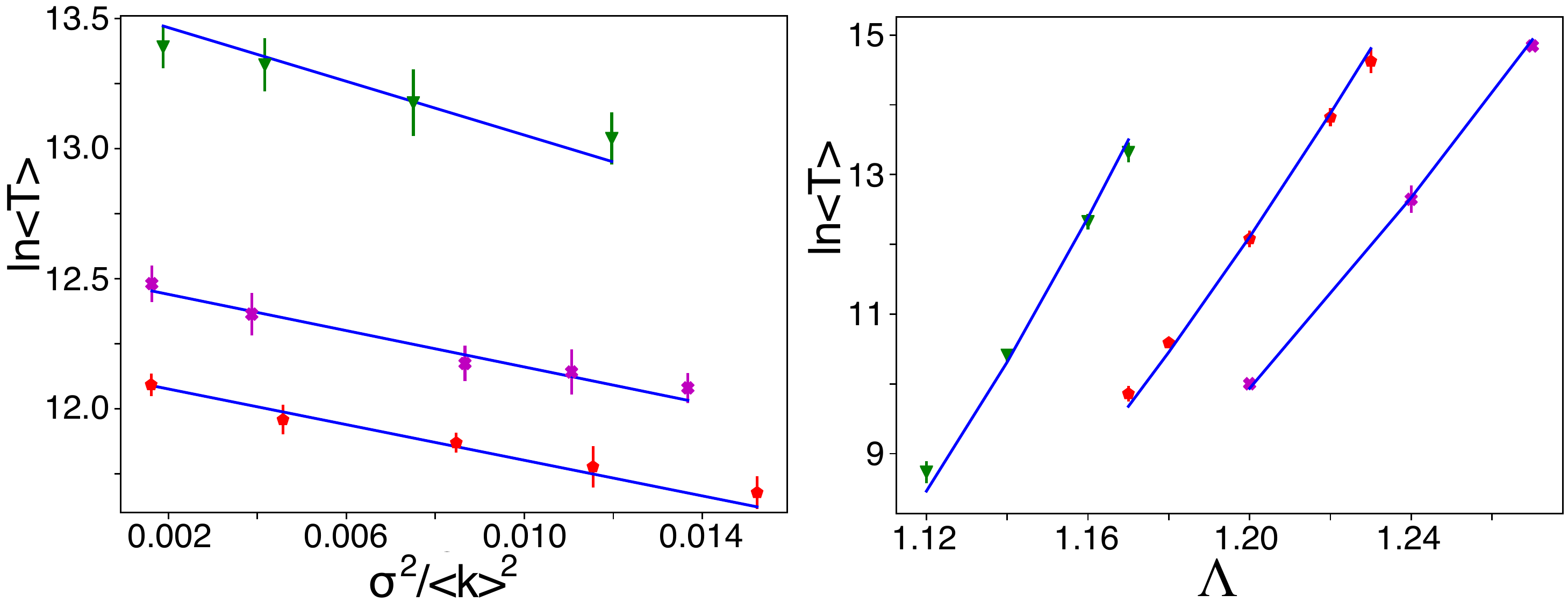}
	\caption{MST versus (left) the degree dispersion and (right) the threshold parameter. The same networks were used as in Fig.~\ref{fig:actioneps}; (left): green, $\Lambda\!=\!1.12$; red, $\Lambda\!=\!1.16$; magenta, $\Lambda\!=\!1.18$}
	\label{fig:actioneps2}
\end{figure}

As was the case for extinction, the action for switching is reduced from the homogeneous network limit by a universal correction, which is a product of the network's CV squared with a model-dependent (though topologically independent) prefactor. As a consequence, the broader the network degree distribution, the more likely switching is to occur between stable magnetization states, given a constant distance to threshold.
Figure \ref{fig:actioneps2} shows a comparison between Eq.~(\ref{eq:spinform}) and Monte-Carlo simulations for the MST in several networks, analogous to Fig.~\ref{fig:actioneps}. As with extinction, the results hold for skewed distributions.

\begin{figure}[!h]
         \center{
	\includegraphics[scale=0.36]{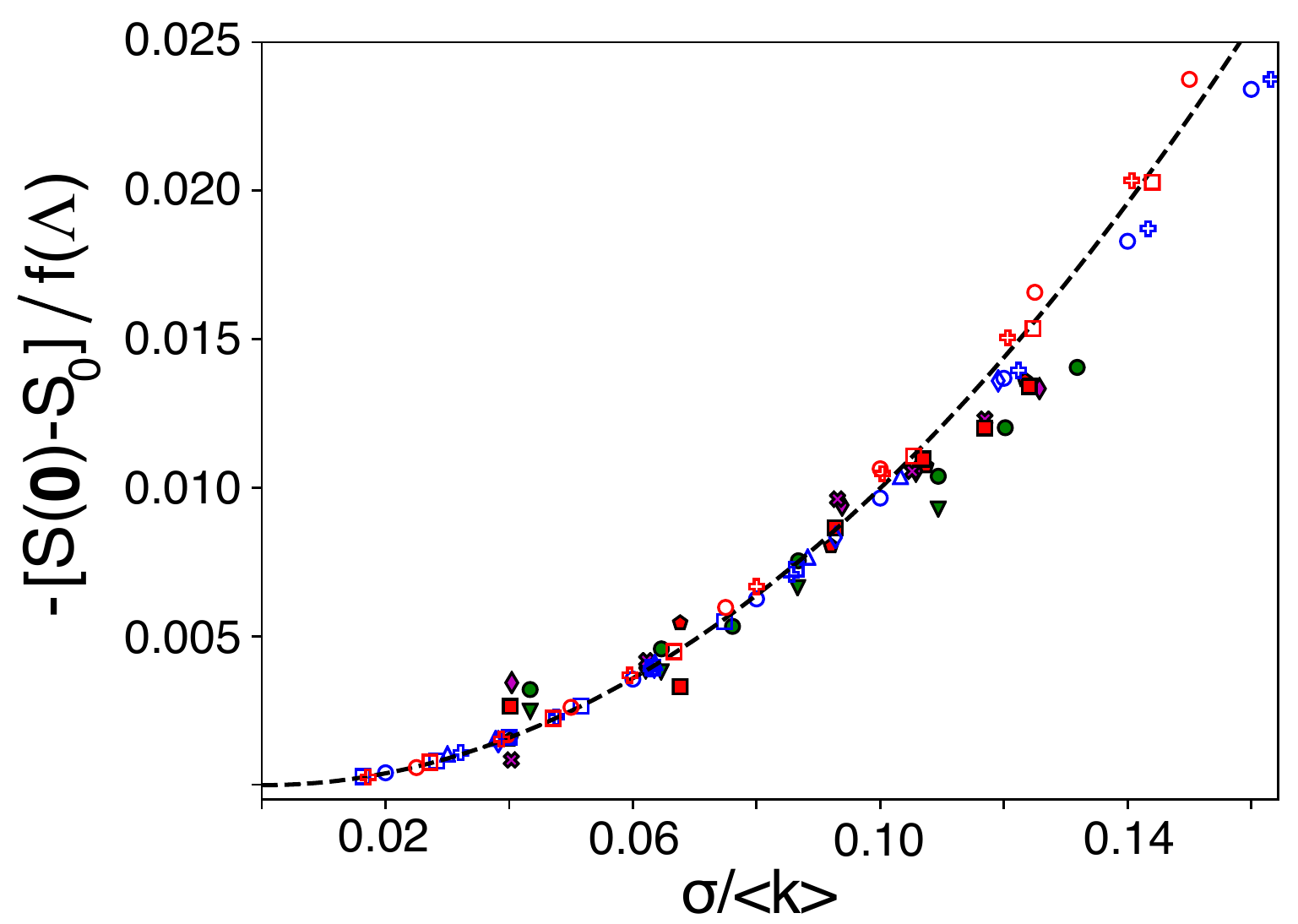}
	\caption{Universal correction to the action for extinction and switching versus the CV; $\Lambda$ ranges from $1.4$ to $3.5$. Solid markers denote network simulations and follow Figs.\ref{fig:actioneps}-\ref{fig:actioneps2}. Numerical computations are shown with open markers for extinction (red) and switching (blue) \cite{SM}. Dashed line is $y=x^2$.}
	\label{fig:collapse}}
\end{figure}

To check the universality of our results, in Fig.~\ref{fig:collapse} we plot the correction $[S(\mathbf{0})-S_{0}]/f(\Lambda)$ versus the CV, and obtain a collapse across all networks and all $\Lambda$, for both models: network simulations and numerical solutions of the Hamilton equations~\cite{SM}. As our analysis exemplifies, if the rate of rare events (on log scale) is normalized by the correct process-dependent factor, $f(\Lambda)$, all networks with the same CV collapse onto the same parabola, given a fixed distance to threshold. Moreover, similar plots and results are shown in the SM for power-law networks and continuous-noise analogs for both processes~\cite{SM}.

To conclude, we employed a novel perturbation theory that utilizes the \textit{extent} of heterogeneity in a network, on two prototypical examples of rare events in networks: extinction in the SIS model of epidemics, and spontaneous magnetization switching in a dynamical spin network. We computed the rate of increase of rare events, and showed that it depends solely on the coefficient of variation (CV) of the network's degree distribution, but is independent of the exact type of network and connectivity matrix. A key insight therein, was to compare different networks with the same distance to threshold, such that deterministic or fluctuation-free stability was held constant, while propensities for noise-induced fluctuations could be isolated. We found that the rate of extinction or switching can be dramatically increased, as long as the CV of the network's degree distribution exceeds $N^{-1/2}$, which is a reasonable assumption for realistic networks. Finally, we have shown that our approach is valid in processes with maintained as well as broken detailed balance, holds across a broad range of network topologies, and generalizes to different noise sources\cite{SM}. Thus, we conjecture that our results are applicable to rare events in a wider range of network processes driven by noise, which include local interactions, and where fluctuations drive a network from a metastable state to an unstable state who merge in a single fixed-point bifurcation\cite{SM}.



We thank Lev Muchnik and Ira B. Schwartz for useful discussions, and Baruch Meerson for critically reading the manuscript. MA was supported through the Israel Science Foundation Grant No. 300/14 and the United States-Israel Binational Science Foundation grant No. 2016-655. JH was supported through the U.S Naval Research Laboratory Karle Fellowship.

\section*{\Large{Supplemental Material}}

\section{\label{sec:SISWKB}SIS Hamiltonian for arbitrary degree distributions}
Following the main text, we first write a master equation for ${\bf n}$, where $n_{k}$ is the number of infected nodes with degree $k$, $N$ is the total number of nodes in the network, $N_k$ is the total number of nodes of degree $k$, and $g(k)=N_k/N$ is the node degree distribution.
Given the annealed network approximation and current state ${\bf n}$, the rate at which $n_{k}$ increases by one is $\lambda k(N_{k}-n_{k})\bar{x}$, where $\bar{x}\equiv\sum_{k'}k'n_{k}'/[N\!\left<k\right>]=\sum_{k'}g_{k'}k'x_{k'}/\!\left<k\right>$, is the fraction of infected neighbors along an edge, and $x_{k}=n_k/N_k$ is the fraction of infected nodes of degree k. Similarly, the rate at which $n_{k}$ is decreased by one is $n_{k}$. We can denote these transitions, compactly, with the notation ${\bf n}\rightarrow{\bf n}+\boldsymbol{1}_{k}$ and ${\bf n}\rightarrow{\bf n}-\boldsymbol{1}_{k}$, respectively.

Consequently, the master equation reads
\begin{align}
\label{eq:SISMaster}
&\frac{\partial P({\bf n},t)}{\partial t}=\sum_{k}\Bigg[\!(n_{k}+1)P({\bf n}+{\bf 1}_{k},t) - n_{k}P({\bf n},t) \;\;+ \nonumber \\
&\lambda(N_{k}-n_{k}+1)k\bar{x}P({\bf n}-{\bf 1}_{k},t) -\lambda(N_{k}-n_{k})k\bar{x}P({\bf n},t) \!\Bigg].\nonumber \\
\end{align}
Now we assume the system has entered a long-lived metastable state, such that $\partial P({\bf n},t)/\partial t\simeq 0$, use the WKB ansatz for the quasi-stationary distribution $P({\bf n})\sim e^{-NS({\bf x})}$, and keep only leading-order terms in $N\gg 1$. This gives rise to a stationary Hamilton-Jacobi equation (where the action has no explicit time dependence), $H({\bf x},\partial_{{\bf x}}S)=0$, with
 \begin{align}
 \label{eq:SISHam}
 H({\bf x},\partial_{{\bf x}}S)=&\sum_{k}g_{k}\Bigg[ \lambda k(1-x_{k})\bar{x}\Big(\!\exp\big\{\partial_{{x}_{k}}S/g_{k}\big\}-1\Big) \nonumber \\
&x_{k}\Big(\!\exp\big\{\!-\partial_{{x}_{k}}S/g_{k}\big\}-1\Big)\Bigg].
\end{align}
The momenta, $\lambda_{k}\equiv\partial_{{x}_{k}}S$, can be usefully redefined as $p_{k}=\lambda_{k}/g_{k}$. With this transformation
$\dot{x}_{k}=\partial_{\lambda_{k}} H({\bf x},{\bf \lambda})=\partial_{p_{k}}H({\bf x},{\bf p})/g_{k}$. Similarly, since $\dot{\lambda}_{k}=-\partial_{x_{k}} H({\bf x},{\bf \lambda})$, we get $\dot{p}_{k}=-\partial_{x_{k}} H({\bf x},{\bf p})/g_{k}$. As a result, the action satisfies
\begin{align}
S({\bf x})=\sum_{k}\int\lambda_{k}dx_{k}=\sum_{k}g_{k}\int p_{k}dx_{k}.
\end{align}

\noindent We note that in Ref.~[21] ``$y_{k}$" is what we call $x_{k}$ in this work.

\section{\label{sec:PertTheory}Finding the optimal path in the SIS model}
In this section we consider a bimodal network with only two degrees $k_1=k_0(1-\epsilon)$ and $k_2=k_0(1+\epsilon)$, where $\langle k\rangle = k_0$ is the mean degree of the network, $\sigma=\epsilon k_0$ is its standard deviation, while $\epsilon\ll 1$. Following the main text, here we find the optimal path to extinction, and the action along it, for such a bimodal network.

To conveniently deal with the Hamiltonian [Eq.~(4) in the main text] in the limit $\epsilon\ll 1$, let us define new variables $u=(x_1-x_2)/2$, $p_u=p_1-p_2$, $w=(x_1+x_2)/2$ and $p_w=p_1+p_2$. This transformation is canonical since the determinant of the Jacobian $\partial(\textbf{Q},\textbf{P})/\partial(\textbf{x},\textbf{p})=1$, where $\textbf{Q}=(u,w)$, $\textbf{P}=(p_u,p_w)$, $\textbf{x}=(x_1,x_2)$, and $\textbf{p}=(p_1,p_2)$. Using the new variables, the path to extinction is a heteroclinic trajectory (or instanton) connecting between the fixed points $[w^*,u^*,0,0]$ and $[0,0,p_w^*,p_u^*]$, where
\begin{eqnarray}\label{fixedpointsSM}
w^*&=&x_0\left[1-(2/\Lambda)\epsilon^2\right],\;\;\;u^*=-(x_0/\Lambda)\epsilon,\\
p_w^*&=&-2\ln\Lambda+[x_0(3\Lambda+1)/\Lambda]\epsilon^2,\;\;\;p_u^*=2x_0\epsilon,\nonumber
\end{eqnarray}
and $x_0=(\Lambda-1)/\Lambda$. Since the transformation of variables is canonical, the action along the path to extinction is given by~[1]
\begin{equation}\label{actionSISSM}
S(\mathbf{0})=\frac{1}{2}\int p_1dx_1+\frac{1}{2}\int p_2dx_2=\frac{1}{2}\int p_wdw+\frac{1}{2}\int p_udu.
\end{equation}

\begin{figure}
	\includegraphics[scale=0.31]{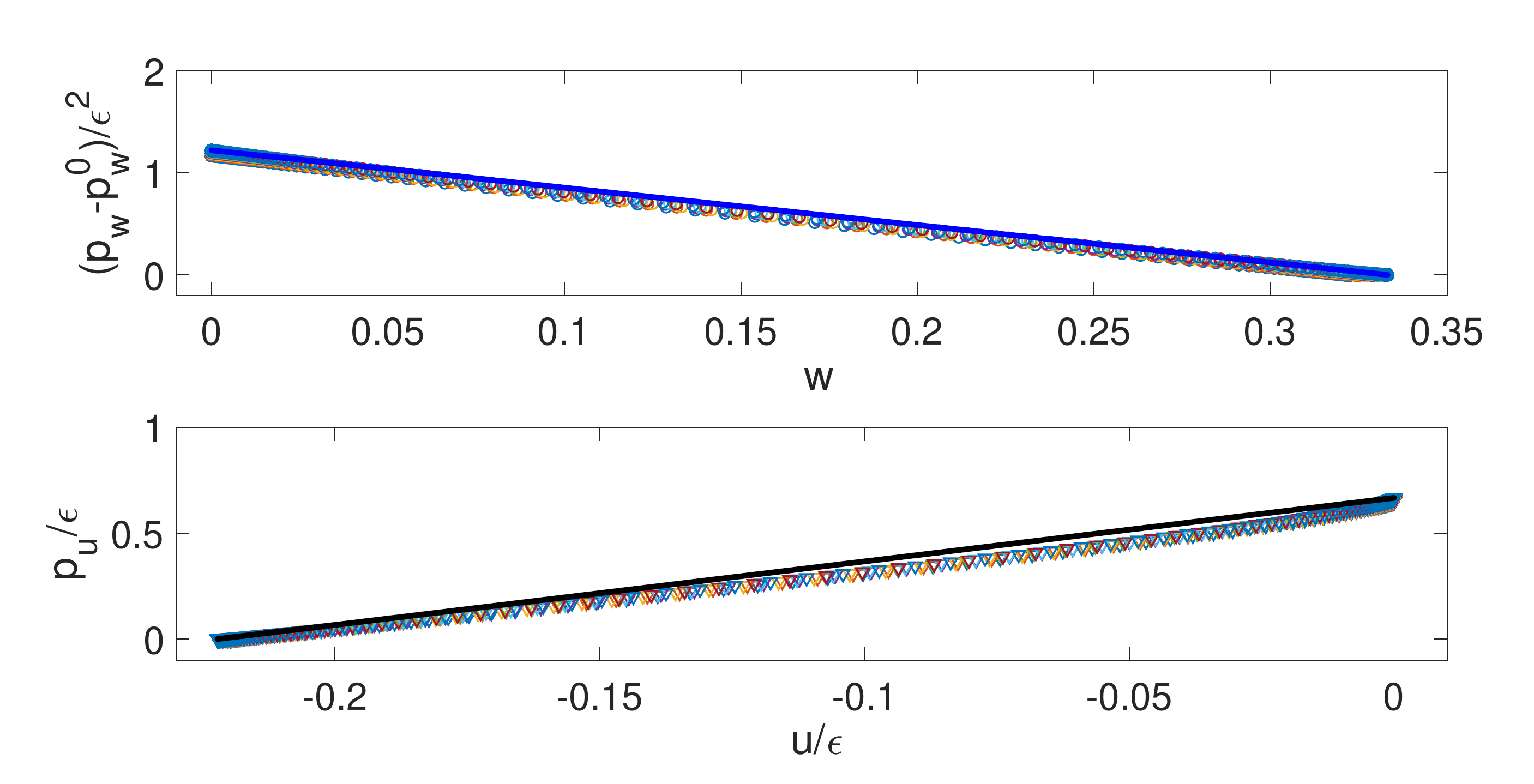}
	\caption{Upper panel: $p_w$ minus the zeroth-order term divided by $\epsilon^2$ as a function of $w$. The symbols are numerics and the line is the analytical result given by Eq.~(\ref{pwSIS}). Lower panel: $p_u/\epsilon$ as a function of $u/\epsilon$. Symbols are numerics and the line is the analytical solution given by Eq.(\ref{puSIS}). Here $\Lambda=2$ and the numerical plots are made for eight different values of $\epsilon$: $0.02,0.04,\dots,0.16$.}
	\label{fig1}
\end{figure}

In the following we find the trajectories $p_w(w)$ and $p_u(u)$, and compute the integral~(\ref{actionSISSM}). We begin by finding $p_w(w)$. Plugging $x_1=w+u$, $x_2=w-u$, $p_1=(p_w+p_u)/2$ and $p_2=(p_w-p_u)/2$ into the Hamiltonian [Eq.~(4) in the main text], and assuming $u$ and $p_u$ scale as ${\cal O}(\epsilon)$, we find in the leading order
\begin{equation}
H(w,p_w,u,p_u)\!=\!2w(e^{p_w/2}-1)[\Lambda(1-w)-e^{-p_w/2}]+{\cal O}(\epsilon^2).
\end{equation}
As a result, we find in the leading order $p_w^{(0)}=-2\ln[\Lambda(1-w)]$. To find the subleading ${\cal O}(\epsilon^2)$ correction, we demand that (i) $p_w$ vanish at $w=w^*$, and (ii) $p_w=p_w^*$ at $w=0$. If we simply interpolate between the two fixed points of $p_w(w)$ by using a linear function of $w$, we get
\begin{equation}\label{pwSIS}
p_w(w)=p_w^{(0)}+\left[3(1-w)-\frac{1+2\Lambda}{\Lambda^2}+\frac{w(3+w)}{\Lambda(1-w)}\right]\epsilon^2.
\end{equation}
One can check a-posteriori that $p_w(0)=p_w^*$ and $p_w(w^*)=0$ up to ${\cal O}(\epsilon^4)$ corrections. In Fig.~\ref{fig1} we numerically verify that Eq.~(\ref{pwSIS}) holds up to ${\cal O}(\epsilon^2)$.
Note, that the numerical solutions of the Hamilton equations, which yield the optimal paths to extinction/switching and the corresponding actions along these paths, were found by using the Iterative Action Minimization Method, see Ref.~[26] for further details. Matlab code is available upon request.

Regarding $p_u(u)$, we notice that both $p_u^*$ and $u^*$ scale as ${\cal O}(\epsilon)$, and thus we expect both $u$ and $p_u$ to scale as ${\cal O}(\epsilon)$ in the entire path. Since the integral over $p_u du$ already scales as ${\cal O}(\epsilon^2)$, it is sufficient to approximate $p_u(u)$ as a straight line connecting $[u^*,0]$ and $[0,p_u^*]$:
\begin{equation}\label{puSIS}
p_u(u)=\frac{2(\Lambda-1)\epsilon}{\Lambda}\left[1+\frac{\Lambda^2 u}{\epsilon(\Lambda-1)}\right],
\end{equation}
which vanishes at $u=u^*$ and equals $p_u^*$ at $u=0$. Again, this choice of path agrees well with numerics, see Fig.~\ref{fig1}.

Finally, performing the integrations in Eq.~(\ref{actionSISSM}) using Eqs.~(\ref{pwSIS}) and (\ref{puSIS}) and keeping terms up to ${\cal O}(\epsilon^2)$, gives
\begin{eqnarray}\label{intwuSIS}
\frac{1}{2}\int p_wdw&\simeq& S_0\!-\!\left[\frac{(\Lambda-1)(3\Lambda^2-10\Lambda-1)}{4\Lambda^3}+\frac{2}{\Lambda}\ln\Lambda\right]\epsilon^2,\nonumber\\
\frac{1}{2}\int p_udu&\simeq& -\frac{(\Lambda-1)^2}{2\Lambda^3}\epsilon^2,
\end{eqnarray}
where $S_0=1/\Lambda+\ln\Lambda-1$ is the action for a degree-homogeneous network ($\epsilon=0$).

\section{Extension of the SIS result to non-symmetric distributions}
Here we generalize Eq.~(9) in the main text to non-symmetric degree distributions.
For any degree distribution, the action along the optimal path is given by
\begin{equation}\label{genactSISSM}
S(\mathbf{0})=\sum_{k=1}^{\infty} g(k) \!\!\int\! p_k dx_k.
\end{equation}
Let us assume a general distribution $g(k)$ centered about $k_0$, with $\sigma\ll k_0$. Thus, it is sufficient to take the sum up to $2k_0$, since the width is much smaller than the mean and $g(2k_0)$ is already  negligible. Denoting by $I_k=\int p_kdx_k$, we have
\begin{eqnarray}\label{nonsysSIS}
\hspace{-5mm}S({\bf 0})&=&\sum_{k=1}^{\infty} g(k)I_k\simeq g(k_0)I_{k_0}\nonumber\\
\hspace{-5mm}&+&\sum_{j=1}^{k_0-1}[g(k_0-j)I_{k_0-j}+g(k_0+j)I_{k_0+j}]\nonumber\\
\hspace{-5mm}&=&g(k_0)I_{k_0}+\sum_{j=1}^{k_0-1}g(k_0-j)[I_{k_0-j}+I_{k_0+j}(1+\delta_j)],\nonumber\\
\end{eqnarray}
where $\delta_j=[g(k_0+j)-g(k_0-j)]/g(k_0-j)$  denotes the deviation from symmetry of the degree distribution. Taylor-expanding $\delta_j$ around $k_0$ up to third order, we find $\delta_j\simeq [g'''(k_0)/(3g(k_0))]j^3$, where prime denotes differentiation with respect to the degree $k$. Evaluating this term at $j=\sigma$, where the distribution has already decayed by a factor of $e$, we find $\delta_j(j=\sigma)\simeq [g'''(k_0)/(3g(k_0))]\sigma^3$. We have evaluated this term for various examples of degree distributions including the Poisson and Gamma distributions, and found in all examples that $\delta_j(j=\sigma)$ is proportional to the distribution's skewness $\gamma_1$. Therefore, for distributions with a small skewness, $\delta_j\sim \gamma_1\ll 1$ for $j={\cal O}(\sigma)$. For smaller $j$, obviously $\delta_j$ is smaller (and again negligible compared to $1$), as we are in the symmetric region of the distribution, while for $j\gg\sigma$, the distribution has already decayed and the terms in the sum are negligible. As a result, we can safely neglect $\delta_j$ in Eq.~(\ref{nonsysSIS}) for all $j$'s, and we recover Eqs.~(8) and (9) in the main text, which were derived for symmetrical distributions.
\section{Finding the optimal path and action in the spin model}
Here we consider the spin model and find the switching path (or instanton) along which the action can be calculated. To do so, we use Hamilton's equations $g(k)\dot{x}_{k}\!=\!\partial H/\partial p_{k}$, where the Hamiltonian is given by Eq.~(10) in the main text. The relevant trajectories $p_{k}(\mathbf{x})$ can be found by equating $-\dot{x}_{k}|_{\mathbf{p}=\mathbf{0}}=\dot{x}_{k}(\mathbf{p})$, where the former represents the deterministic trajectory. By doing so, we obtain the following equations for $p_k(\mathbf{x})$:
\begin{eqnarray}
&&e^{2p_k}(1\!-\!x_k)\!\left(\frac{1}{1\!+\!e^{-2\lambda k\bar{x}}}\right)-e^{-2p_k}(1\!+\!x_k)\!\left(\frac{1}{1\!+\!e^{2\lambda k\bar{x}}}\right)\nonumber\\
&&=(1+x_k)\!\left(\frac{1}{1\!+\!e^{2\lambda k\bar{x}}}\right)-(1-x_k)\left(\frac{1}{1\!+\!e^{-2\lambda k \bar{x}}}\right).
\end{eqnarray}
After some algebra, we find a solution
\begin{eqnarray}
p_k(\mathbf{x})=\frac{1}{2}\ln\left(\frac{1+x_k}{1-x_k}\right)-\lambda k\bar{x},
\end{eqnarray}
which leads to the action [Eq.~(11) in the main text].

In order to approximate the action in the limit of $\left<|\epsilon|\right>=\sigma/k_0\!\ll\!1$ we need to first evaluate $x_{k}^*$ and $\bar{x}^*$ in that limit. Using the Hamiltonian [Eq.~(10) in the main text], the deterministic rate equations (when $\mathbf{p}\!=\!\mathbf{0}$) have fixed points $x_{k}^{*}$ which satisfy the following transcendental equations:
$x_{k}^{*}\!=\!\tanh\{\lambda k \bar{x}^{*}\}$ [36,37]. If we assume that $\bar{x}^{*}$ takes the form $\bar{x}^{*}\!=\!x_{0}+\left<\epsilon^{2}\right>x_{1}$, where $x_{0}$ is the positive solution of $x_{0}\!=\!\tanh\{\Lambda x_{0}\}$, then
\begin{align}
x_{k}^{*}=&x_{0}+\Lambda x_{0}(1-x_{0}^{2})\epsilon-\Lambda^{2}x_{0}^{3}(1-x_{0}^{2})\epsilon^{2} \nonumber \\
&+\Lambda(x_{1}-x_{0})(1-x_{0}^{2})\left<\epsilon^{2}\right> +\mathcal{O}(|\epsilon|^{3}).
\label{eq:Xkstar}
\end{align}
Substituting Eq.~(\ref{eq:Xkstar}) into the definition of $\bar{x}^{*}$ we find
\begin{equation}\label{spinmbar}
\bar{x}^{*}=x_0\left[1-\frac{\Lambda^{2}x_0^2(1-x_0^2)}{1-(1-x_0^2)\Lambda}\left<\epsilon^{2}\right>\right],
\end{equation}
where $x_1$ in Eq.~(\ref{eq:Xkstar}) satisfies $x_1=-\Lambda^{2}x_0^3(1-x_0^2)/[1-(1-x_0^2)\Lambda]$. Plugging Eqs.~(\ref{eq:Xkstar}) and (\ref{spinmbar}) into the action [Eq.~(11) in the main text] yields the final result for the mean switching time
\begin{equation}\label{eq:spinformSM}
T\sim e^{NS(\mathbf{0})},\;\;\;S(\mathbf{0})=S_{0}-\frac{\Lambda x_0^2}{2}\left[1-\Lambda(1- x_0^2)\right]\frac{\sigma^{2}}{\left<k\right>^{2}},
\end{equation}
where $S_{0}=-(1/2)\left[\ln\left(1-x_0^2\right)+\Lambda x_0^2\right]>0$. This result coincides with Eq.~(12) in the main text.

\section{Breaking detailed balance in the spin model}
Here we generalize our results for the spin model \textit{in the absence of detailed balance}.
A simple way to break detailed balance is to add a spontaneous transition with rate $f$. Namely, we assume that each node flips spin at a stochastic rate, $f+[1+\exp{\!\{\lambda\Delta E_{i}\}}]^{-1}$. In the presence of this spontaneous flipping process, the Hamiltonian [Eq.~(10) in the main text] becomes:
\begin{align}
\!\!H({\bf x},{\bf p})\!=\!\sum_{k}g_{k}&\!\Bigg[\frac{1}{2}(1\!-\!x_{k})(e^{2p_{k}}\!-\!1)\!\Bigg(\!\frac{1}{1\!+\!e^{-2\lambda k\bar{x}}}+f\!\Bigg)\nonumber \\
&\!\!\!\!\!+\frac{1}{2}(1\!+\!x_{k})(e^{-2p_{k}}\!-\!1)\!\Bigg(\!\frac{1}{1\!+\!e^{2\lambda k\bar{x}}}+f\!\Bigg)\!\Bigg]\!,
\label{eq:HamiltonianDBB}
\end{align}
which can be derived in exactly the same way as above for the SIS model (see Ref.~[37]). It is straightforward to show that the pitchfork bifurcation now occurs when $\lambda\left<k^{2}\right>\!/\!\left<k\right> -1-2f\!=\!0$.

\begin{figure}[ht]
\includegraphics[scale=0.48]{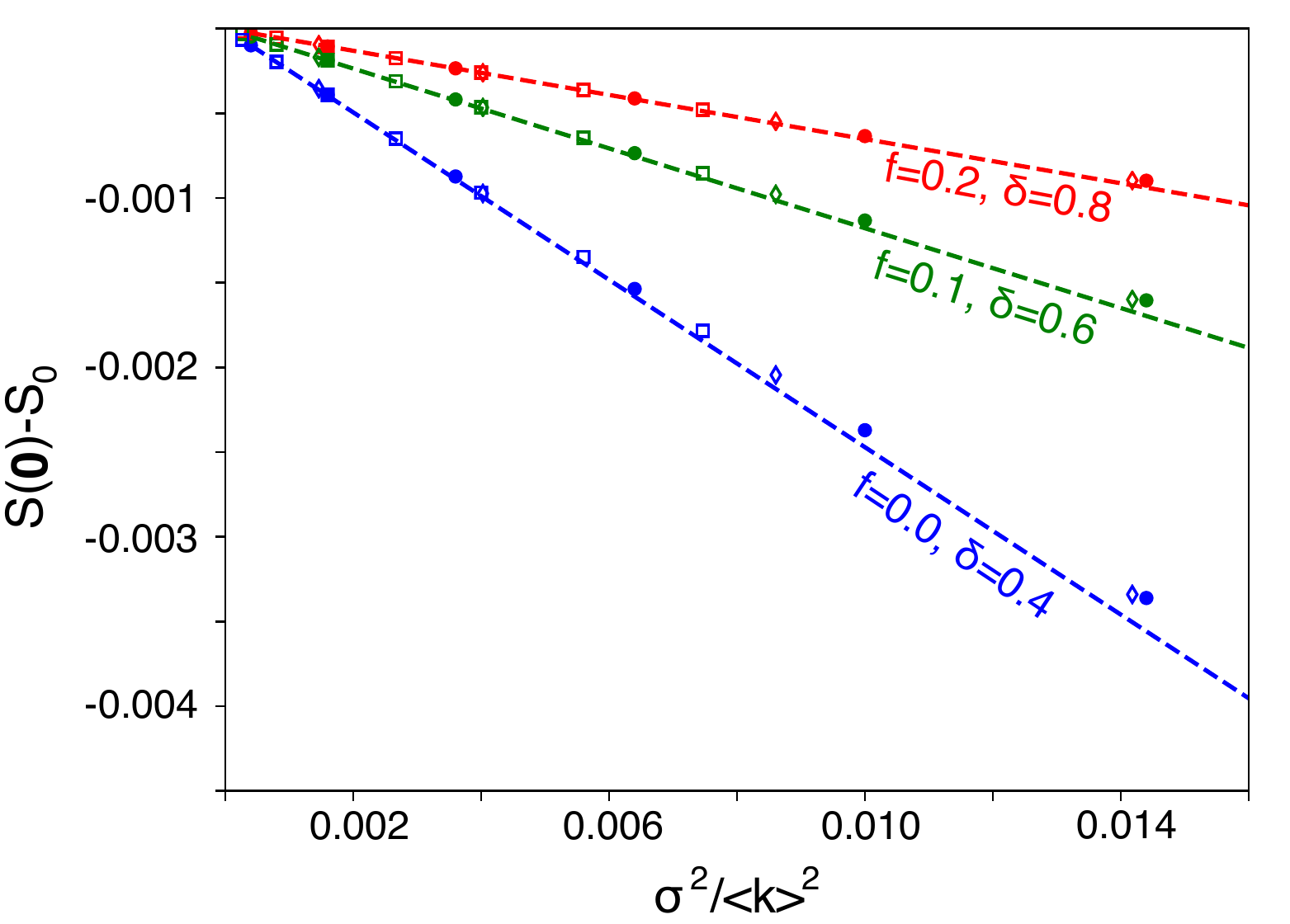}
\caption{Change in the action, relative to the homogeneous network limit, versus the coefficient of variation squared; bimodal distributions with $\left<k\right>\!=\!50$ (circles), uniform distributions with $\left<k\right>\!=\!50$ (squares), and Gamma distributions with $\left<k\right>\!=\!108.5$  (diamonds).}
\label{fig:actionepsDBB}
\end{figure}

The action for switching can be computed from
\begin{align}
S({\bf 0})=\sum_{k}g_{k}\!\int_{x_{k}^{*}}^{0}p_{k}dx_{k},
\end{align}
where $g_{k}\dot{x}_{k}\!=\!\partial H/\partial p_{k}$ and $g_{k}\dot{p}_{k}\!=-\partial H/\partial x_{k}$. We solve this system numerically for several networks and values of $f$; the results are shown in Fig.~\ref{fig:actionepsDBB}. In order to keep the distance to bifurcation constant across all networks used, we define $\lambda\!=\![1+\delta]\left<k\right>\!/\!\left<k^{2}\right>$. Therefore, all three series in Fig.~\ref{fig:actionepsDBB} have the same distance to bifurcation, $\lambda\left<k^{2}\right>\!/\!\left<k\right> -1-2f$.\\

Our numerical results indicate that, even in the absence of detailed balance, the correction to the action across {\it all networks} collapses to the same expression
\begin{align}
S({\bf 0})=S_{0} -f_{S}(\delta,f)\frac{\sigma^{2}}{\left<k\right>^{2}},
\end{align}
and hence, our main result is preserved.
Note however, that $f_{S}(\delta,f)$ is no longer a function, only, of the distance to bifurcation -- otherwise all three series would collapse to the same correction. This more general function could be calculated, i.e., with the general procedure used for extinction in the SIS model, without assuming detailed balance; see main text and Sec. SM-\ref{sec:PertTheory}.

\section{Parameters for Fig.~4 in main text}
Here we describe in detail the results shown in Fig.~4 in the main text. The network simulations for this figure were taken from Fig.~2 (left) and Fig.~3 (left) in the main text. Furthermore, Fig.~4 in the main text includes numerical solutions of Hamilton's equations. In red we show the numerical results for extinction; circles are bimodal distributions with $\left<k\right>\!=\!40$ and $\Lambda\!=\!3.5$, squares are uniform distributions with $\left<k\right>\!=\!30$ and $\Lambda\!=\!2.0$, crosses are generalized Gaussian distributions with exponent $1.0$, $\left<k\right>\!=\!35$, and $\Lambda\!=\!3.0$. In blue, we show numerical results for switching; circles are bimodal distributions with $\left<k\right>\!=\!50$ and $\Lambda\!=\!1.4$, squares are uniform distributions with $\left<k\right>\!=\!50$ and $\Lambda\!=\!1.4$, diamonds are Gamma distributions with $\left<k\right>\!=\!50$ and $\Lambda\!=\!1.4$, triangles are Gaussian distributions with $\left<k\right>\!=\!108.5$ and $\Lambda\!=\!1.7$, and crosses are generalized Gaussian distributions with exponent $0.75$, $\left<k\right>\!=\!35$, and $\Lambda\!=\!1.7$. Note that given these parameters, each degree distribution has a single parameter which can be varied to change the coefficient of variation. Finally, the dashed-line in Fig.~4 in the main text is the theoretical prediction $-[S({\bf 0})-S_0]/f(\Lambda)=\sigma^2/\langle k\rangle^2$.
\\~\\
Here and in Figs.~2 and 3 in the main text the simulations on networks were performed using Monte-Carlo simulations implemented according to the Gillepsie's algorithm in continuous time. Namely, for each node in a network, there is an exponentially distributed time to make a transition to another state. For example, a transition of a susceptible node to infected occurs at a rate $\lambda$ times the number of infected neighbors. Noise comes from the fact that the time is not deterministic, but is a stochastic variable. C++ code is available upon request.

\section{Continuous models with continuous noise}
In the main text, we deal with discrete states on the nodes. However, our results are qualitatively the same for continuous states with continuous noise, and an analogous perturbation-theory in $\sigma\!/\!\left<k\right>$ can be developed. In particular the network action for extinction/switching takes the form, $S\big(\Lambda,\frac{\sigma}{\left<k\right>}\big)\approx S\big(\Lambda,0\big)-f(\Lambda)\frac{\sigma^{2}}{\left<k\right>^{2}}$.

Let us consider the following Langevin system
\begin{equation}\label{Continuous}
\dot{x}_{i}=\mathcal{M}_{i}(\mathbf{x};\lambda) +\xi_{i}(t),
\end{equation}
where $\mathcal{M}_{i}(\mathbf{x};\lambda)$ is the {\it mean-field} dynamics for node $i$, and $\xi_{i}(t)$ is independent and identically distributed Gaussian white noise (GWN), $\left<\xi_{i}(t)\xi_{j}(t')\right>=2D\delta_{ij}\delta(t-t').$ The mean-field dynamics correspond to $\mathbf{p}\rightarrow\mathbf{0}$ in Hamilton's equations, or
\begin{align}
\mathcal{M}_{i}^{(\text{SIS})}=\;&\lambda k_{i}(1-x_{i})\sum_{j}\!\frac{k_{j}x_{j}}{N\left<k\right>}-x_{i}, \\
\mathcal{M}_{i}^{(\text{spin})}=\;&\tanh\!\Bigg\{\lambda k_{i}\sum_{j}\!\frac{k_{j}x_{j}}{N\left<k\right>}\Bigg\}-x_{i}
\end{align}
for the SIS and spin models, respectively.

Similar to the main text, the quasi-stationary probability distributions have a WKB form when $D\ll1$, $P(\mathbf{x})\!\sim\!\exp\{-NS(\mathbf{x})/[2D]\}$. Hamilton's equations are straightforward to derive (see for instance E. Forgoston and R. O. Moore, SIAM Rev. {\bf 60}(4), 969 (2018)), and represent an application of classical large-deviation theory for dynamical systems perturbed by GWN. The procedure for deriving Hamilton's equations is essentially the same as in the main-text and Sec. SM-\ref{sec:SISWKB}, except the master equation, e.g. Eq.(\ref{eq:SISMaster}), is replaced by a Fokker-Planck equation for Eq.(\ref{Continuous}).

 Given that we expect nodes with the same degree to have synchronized dynamics during a large fluctuation (i.e., trading the node subscript $i$ for the degree subscript $k$), we find for the SIS model
\begin{align}
\label{eq:Cont1}
\dot{x}_{k}=&\;\lambda k(1-x_{k})\bar{x}-x_{k} + p_{k},\\
\dot{p}_{k}=&\;p_{k}\big[\lambda k \bar{x}+1\big]-\lambda k\!\sum_{k'}\!\frac{k'g_{k'}p_{k'}}{\left<k\right>}(1-x_{k}'),
\end{align}
and for the spin model
\begin{align}
\dot{x}_{k}=&\;\tanh\!\big\{\lambda k\bar{x}\big\}-x_{k} + p_{k},\\
\label{eq:Cont2}
\dot{p}_{k}=&\;p_{k} -\lambda k \!\sum_{k'}\!\frac{k'g_{k'}p_{k'}}{\left<k\right>}\text{sech}^{2}\!\big\{\lambda k'\bar{x}\big\},
\end{align}
where, as above, $\bar{x}\!\equiv\!\sum_{k'}\!k'x_{k'}/[N\!\left<k\right>]$ and $S\!=\!\sum_{k}g_{k}\int p_{k}dx_{k}$.

\begin{figure}[ht]
\includegraphics[scale=0.26]{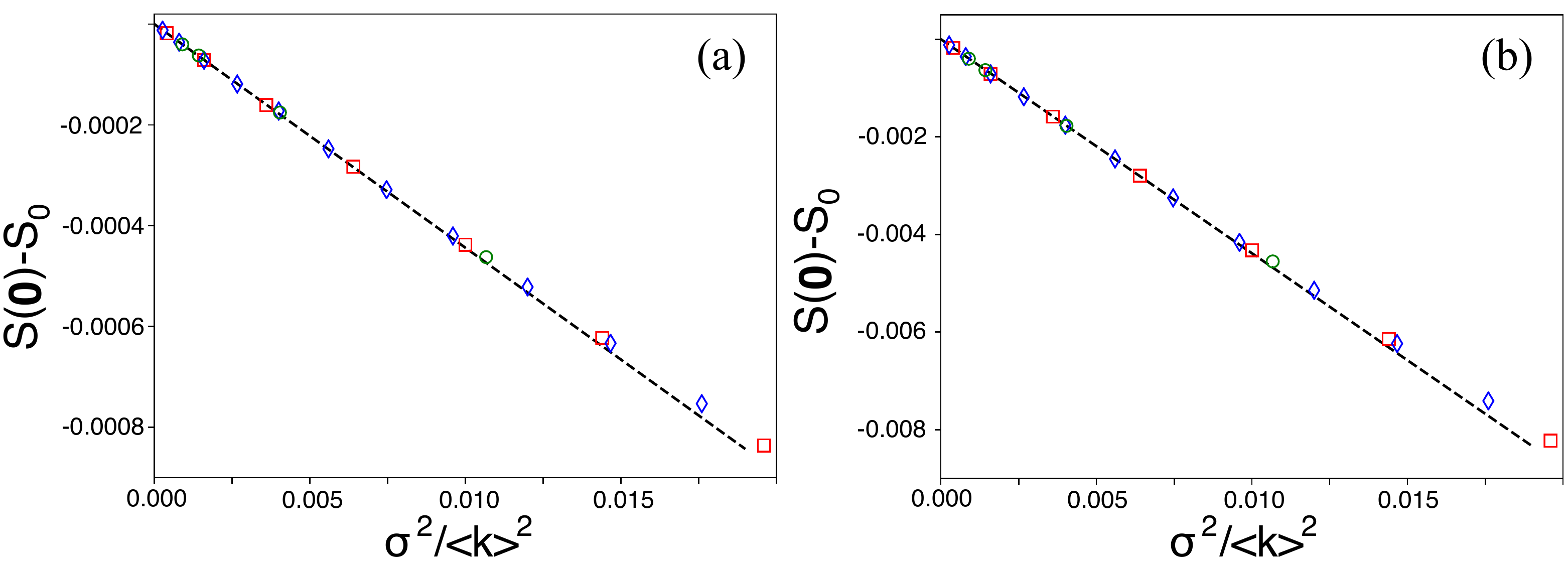}
\caption{Change in the action for continuous noise models, relative to the homogeneous network limit, versus the coefficient of variation squared; bimodal distributions with $\left<k\right>\!=\!50$ (squares), uniform distributions with $\left<k\right>\!=\!50$ (diamonds), and Gamma distributions with $\left<k\right>\!=\!108.5$  (circles). (a) SIS model, $\Lambda\!=\!1.5$. (b) spin model, $\Lambda\!=\!1.5$. Actions were computed from numerical solutions of Eqs.~(\ref{eq:Cont1}-\ref{eq:Cont2}). The dashed lines are the theoretical predictions.}
\label{fig:Contin}
\end{figure}

Figure \ref{fig:Contin} shows the change in the action from the homogeneous network limit, for both processes, as a function of $\sigma^{2}\!/\!\left<k\right>^{2}$ for three different examples of degree distributions.  The results are consistent with those presented in the main text.

\section{Power-law degree distributions}
\begin{figure}[ht]
\includegraphics[scale=0.52]{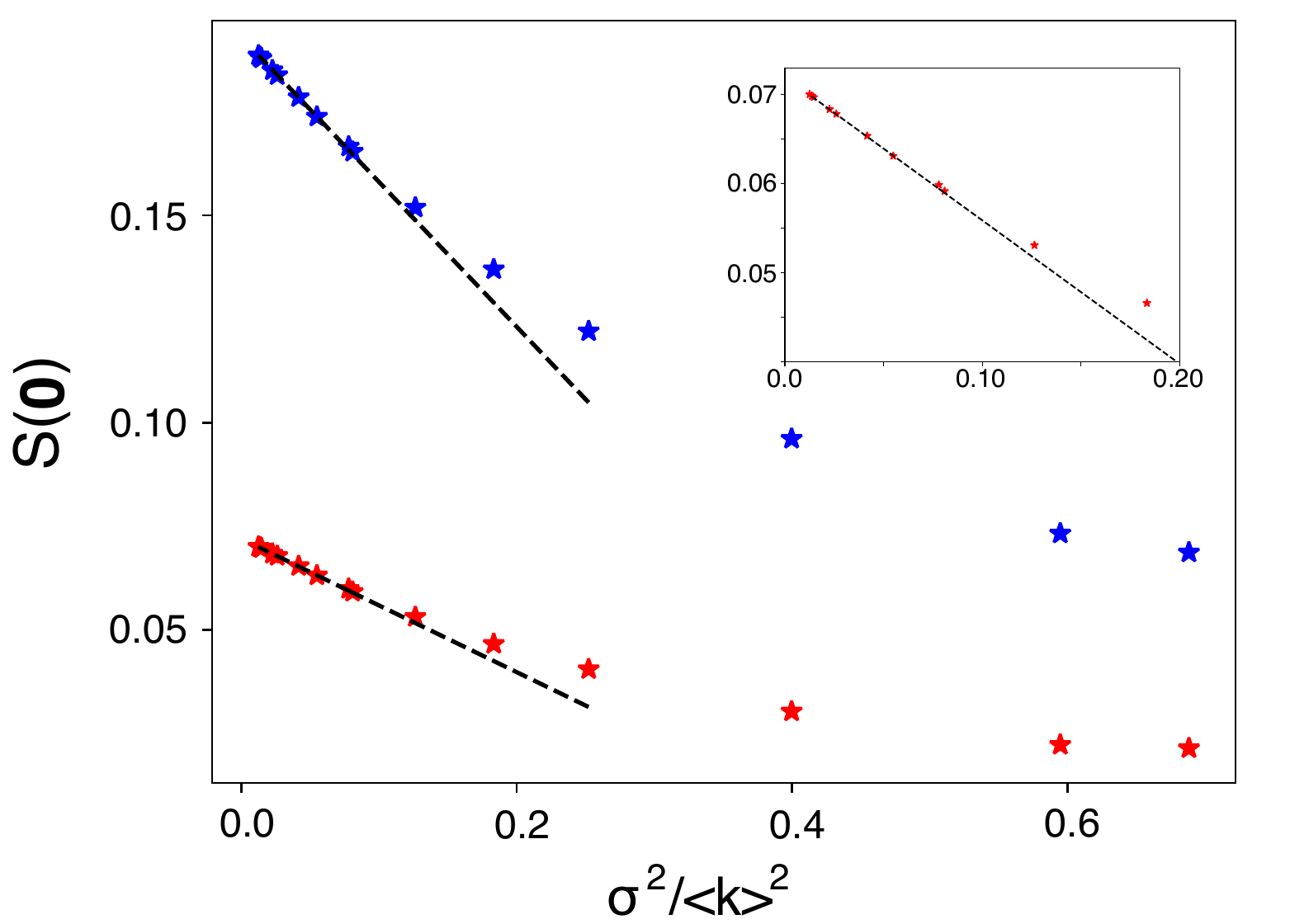}
\caption{Action for extinction in power-law networks with degree distributions $g(k,s)\!=\!k^{-s}\!/\!\sum_{k'=20}^{500}k'^{-s}$. The degree exponent, $s$, ranges from $s\!=\!10,9.5,...,2.5$; $\Lambda\!=\!1.5$ (red) and $\Lambda\!=\!2.0$ (blue). The inset shows $\Lambda\!=\!1.5$ for $s\!>\!4.0$. Actions were computed from numerical solutions of the Hamilton's equations for the SIS model, that can be derived from Eq.~(\ref{eq:SISHam}). Note: for large $s$, points become very close, making them harder to distinguish. The dashed lines are the theoretical predictions.}
\label{fig:PowerLaw}
\end{figure}

In the main-text we primarily discuss networks whose degree-distributions are centralized around a mean, with an approximately symmetric pattern of dispersion. Nevertheless, our quantitative results turn out to also hold for power-law networks with relatively small coefficients of variation (e.g. degree exponents greater than four). Moreover, our qualitative result: degree dispersion increases the rate of rare events when comparing networks with constant distances to threshold, holds for power-law networks with even smaller degree exponents. Figure \ref{fig:PowerLaw} shows the action for extinction for power-law networks with degree distributions $g(k,s)\!=\!k^{-s}\!/\!\sum_{k'=20}^{500}k'^{-s}$. The degree exponent ranges from $s\!=\!10,9.5,...,2.5$, where $\Lambda\!=\!1.5$ (red) and $\Lambda\!=\!2.0$ (blue). The dashed line shows the predicted scaling, $S(\mathbf{0},s)\!\approx\! S(\mathbf{0},s\!\rightarrow\!\infty)-f_{E}(\Lambda)\frac{\sigma^{2}}{\left<k\right>^{2}}$ (Eq.(7) and Eq.(9) in the main text), which agrees well with numerics for $s\!>\!5$. For reference, a power-law network with $s\!=\!5$ has a variance of $\sigma^{2}\!/\!\left<k\right>^{2}\!\approx\!0.12$.

\section{Generality of our results}
In this section we briefly discuss the generality of our results. We have shown that the barrier for extinction/switching, given in the form of a cumulative action obtained by integrating over a trajectory between the deterministically stable and unstable fixed points, \textit{decreases} as the heterogeneity of the network, $\epsilon\!=\!\sigma\!/\!\left<k\right>$, is \textit{increased}. Specifically, we have demonstrated the following functional dependence $S\big(\Lambda,\epsilon\big)\approx S\big(\Lambda,0\big)-f(\Lambda)\epsilon^2$, for both the SIS model of epidemics and a model of spontaneous magnetization flipping, where $f(\Lambda)$ depends on the local microscopic dynamics, but is independent on the network topology. That is, as long as the heterogeneity parameter $\epsilon$ is fixed, we have shown that the network topology affects the mean escape time in a universal manner, regardless of the degree distribution of the network. Moreover, the dependence on heterogeneity holds for distinct types of rare events: extinction and switching.

It is our conjecture, that any model that satisfies the following generic conditions will demonstrate similar quantitative features:

\begin{itemize}
\item{At the microscopic level, the model should include one-body and two-body interactions, where the latter are due to interactions between each node and its neighbors. These microscopic dynamics determine the specific nature of the function $f(\Lambda)$.}
\item{The microscopic dynamics should give rise at the deterministic level to a nontrivial stable state and an adjacent unstable state which is either an absorbing state, or it is accompanied by an additional (target) stable fixed point. Such states should be fixed-points of a mean-field dynamics. The mean-field description should entail sets of differential equations in time for the density, or set of densities, describing the average state of nodes with degree $k$ (or eigenvector centrality).}
\item{The model should include a tuning parameter $\Lambda$, which when approaches some $\Lambda_c$,  the stable fixed point(s) at the deterministic level merge with the unstable fixed point, and the system becomes monostable. That is, the system can undergo, \textit{e.g.}, a transcritical, a pitchfork or a saddle-node bifurcation, depending on the scenario at hand. When noise is accounted for, the former case typically corresponds to an escape from a metastable state to an absorbing state (e.g., extinction), while the latter cases typically correspond to switching between two metastable states separated by a saddle point.}
\end{itemize}
\noindent Note: the state space and noise can be continuous or discrete.

Finally, while we have considered two prototypical examples of extinction and switching, we expect our results to hold for wide variety of additional models which satisfy the conditions specified above. Examples include population dynamics models (or equivalent) with an Allee effect, other models of epidemics such as the SIRS models, and voter models on networks with hysteresis. On the other hand, models of evolutionary game theory on networks and generalized contagion models, which include more complicated bifurcation scenarios, are expected to (possibly) display a different dependence on $\epsilon$ in the action, as the network heterogeneity is increased.

\end{document}